\documentclass[a4paper]{jpconf}
\usepackage{graphicx}
\usepackage{siunitx}
\PassOptionsToPackage{hyphens}{url}\usepackage[hidelinks]{hyperref}
\begin{document}

\begin{center}
{Preprint accepted at CISBAT 2023 - The Built Environment in Transition, Hybrid International Conference, EPFL, Lausanne, Switzerland, 13-15 September 2023}
\end{center}

\title{Cozie Apple: An iOS mobile and smartwatch application for environmental quality satisfaction and physiological data collection}

\author{Federico Tartarini$^{1,2}$, Mario Frei$^3$, Stefano Schiavon$^4$, Yun Xuan Chua$^3$, Clayton Miller$^{3,*}$}
\address{$^1$ SinBerBEST, Berkeley Education Alliance for Research in Singapore, Singapore}
\address{$^2$ Heat and Health Research Incubator, Faculty of Medicine and Health, The University of Sydney, Australia}
\address{$^3$ College of Design and Engineering, National University of Singapore (NUS), Singapore}
\address{$^4$ Center for the Built Environment (CBE), University of California, Berkeley, CA, USA}

\ead{$^{*}$clayton@nus.edu.sg}

\begin{abstract}
    Collecting feedback from people in indoor and outdoor environments is traditionally challenging and complex to achieve in a reliable, longitudinal, and non-intrusive way.
    This paper introduces Cozie Apple, an open-source mobile and smartwatch application for iOS devices.
    This platform allows people to complete a watch-based micro-survey and provide real-time feedback about environmental conditions via their Apple Watch.
    It leverages the inbuilt sensors of the smartwatch to collect physiological (e.g., heart rate, activity) and environmental (sound level) data.
    This paper outlines data collected from 48 research participants who used the platform to report perceptions of urban-scale environmental comfort (noise and thermal) and contextual factors such as who they were with and what activity they were doing. 
    The results of 2,400 micro-surveys across various urban settings are illustrated in this paper, showing the variability of noise-related distractions, thermal comfort, and associated context.
    The results show that participants experienced at least \emph{a little} noise distraction 58\% of the time, with people \emph{talking} being the most common reason (46\%).
    This effort is novel due to its focus on spatial and temporal scalability and the collection of noise, distraction, and associated contextual information.
    These data set the stage for larger deployments, deeper analysis, and more helpful prediction models toward better understanding the occupants' needs and perceptions.
    These innovations could result in real-time control signals to building systems or nudges for people to change their behavior.
\end{abstract}

\section{Introduction}\label{ch:introduction}
Most buildings cannot learn and adjust their operations based on occupants' feedback, preferences, and needs. 
Instead, environmental sensors (e.g., illuminance, thermostats, occupancy, and CO\textsubscript{2}) are designed to control lighting, heating, ventilation, and cooling systems.
As a result, occupants are limited to adjusting what they can control, for example, their clothing~\cite{Tartarini2018a}, or can raise complaints~\cite{Parkinson2023-yk}.
Noise and its effect on productivity and mental health is also a significant issue in the built environment~\cite{Kim2013-gs,Bergefurt2023-fm}.
According to a survey conducted on over 90,000 individuals in approximately 900 office buildings, the primary sources of dissatisfaction among occupants were sound privacy, with 54\% of respondents expressing dissatisfaction, followed by temperature at 39\%, and noise level at 34\%~\cite{Graham2021-en}.

Researchers and practitioners have been installing increasing numbers of sensors in buildings and urban environments to characterize human exposure.
However, as previously noted, sensors' data are rarely coupled with occupants' feedback~\cite{Li2020b}. 
In addition, researchers are moving towards using personal comfort models, also referred to as \emph{occupant-centric control}. 
Personal thermal comfort modeling is a new approach that aims to predict the thermal comfort responses of each person instead of aggregating data from a large population~\cite{Quintana2023, Tartarini2022}. 
Personal comfort models can be based on environmental measurements (e.g., air temperature, location, relative humidity), occupant feedback (e.g., online voting), occupant behavior (e.g., thermostat setpoints), and physiological parameters (e.g., skin temperature, heart rate)~\cite{Kim2018a}. 
In addition, acoustic comfort continues to be overlooked, despite some efforts made to monitor noise pollution levels in urban environments. 
However, these initiatives fail to capture the subjective aspect of whether the sound levels actually act as distractions~\cite{Zipf2020-br,Alias2019-le}.
These challenges face the issue that collecting data from people is not trivial and poses several challenges.
These include but are not limited to obtaining written consent, handling confidential information, reliably delivering the surveys, and finding the optimal balance between the quality and quantity of data collected.
Long surveys delivered frequently often lead to survey fatigue, increasing the likelihood of participants dropping out from the study and collecting unbalanced datasets~\cite{Kim2019}.

\subsection{Development of an open-source interface for watch-based data collection}
To overcome all the issues mentioned above, the open-source Cozie smartwatch platform was developed to change how people can provide feedback about their environment and how investigators collect physiological data.
Many people are already wearing a smartwatch daily; hence Cozie users are not required to wear additional sensors or devices (e.g., heart rate monitor, pedometer).
With a smartwatch-based platform, occupants no longer need to use either pen and paper, phone, or web applications to provide feedback, as done in previous studies~\cite{Sood2020-vz}.
This method reduces inefficiencies and downtime, like getting the phone, unlocking it, and loading the application.
This method enhances and improves Ecological Momentary Assessments (EMA), a type of experience sampling method used in psychology~\cite{Shiffman}.
Ecological Momentary Assessments aim to take measurements and ask for real-time subjective assessments when a defined behavior occurs.

The Cozie platform was originally developed for the Fitbit smartwatch ecosystem~\cite{Jayathissa2019a}.
This paper focuses on subsequent Cozie's development and deployment for the Apple iOS ecosystem.
Cozie Apple extends and improves the capabilities of the Fitbit version, allowing researchers to deliver right-here-right-now surveys to a large group of users without compromising data quality.
The companion app on the phone can be used to complete eligibility and onboarding surveys and the signed consent form, making it scalable to the large community of Apple Watch users. 
The platform also allows push notifications and reminders to be sent to participants, and users can access data in real time.
Additionally, the framework uses the iOS HealthKit API to access physiological and environmental data with user consent. 
The code for the project is open-sourced, and released official documentation and video tutorials can be accessed at \emph{\href{https://cozie-apple.app/}{\url{cozie-apple.app}}}.
This manuscript outlines the results of an urban-scale deployment to collect data in a longitudinal field study to investigate people's perceptions of noise and distraction in the urban context. 
This deployment aims to capture whether a person experiences distractions due to noise and, if they exist, what contextual dimensions are co-occurring~\cite{Miller2022-dy}. 
These contextual elements include whether the person is wearing earphones, if thermal comfort issues arise simultaneously, what type of space they are in, whether they are with others, and what type of activity the person is doing.

\section{Methodology}
Cozie comprises an iOS App (companion application) and a WatchOS App (smartwatch application) built using Swift.
It is backward compatible with iOS 14 and Apple Watches Series 3 or newer.
The application can be downloaded from the Apple Store or deployed using the open-source code base that is at \emph{\href{https://github.com/cozie-app/cozie-apple}{github.com/cozie-app/cozie-apple}}.
Researchers can, therefore, use the open-source code to deploy their own applications to enable the customization of the application based on specific research objectives and requirements.
Cozie offers several functionalities which investigators can modify and edit to suit their needs better.
Changes must be implemented directly in the source code, and a new version of the application needs to be deployed each time a change has to be made.

\begin{figure}[b!]%[htbp]
     \centering
     \includegraphics[width=\textwidth]{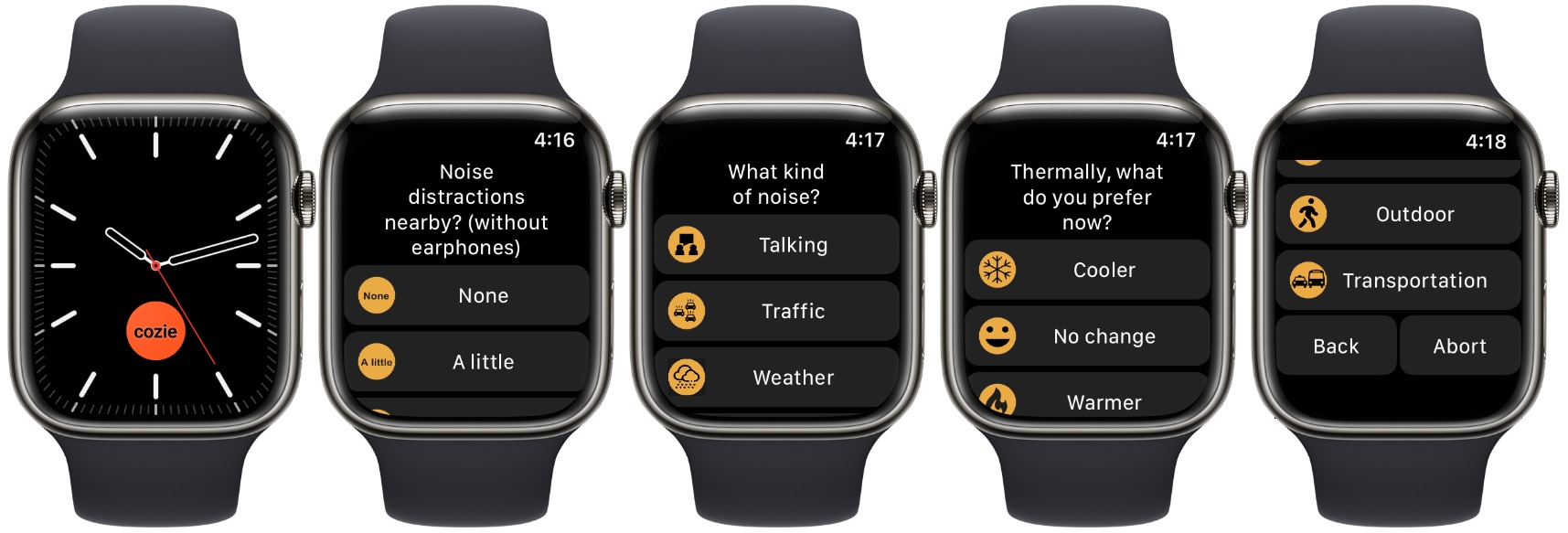}
     \caption{Example of an Apple Watch clock face with the Cozie home screen icon (far left watch face) and examples of several watch survey questions used in the experimental deployment.}
     \label{fig:cozie_flow_and_complications}
\end{figure}

An example of some of the questions that can be added to the smartwatch survey is shown in Figure~\ref{fig:cozie_flow_and_complications}.
The \emph{Back} and \emph{Abort} buttons at the end of each survey question allow users to edit their answers or to stop and discard the complete survey, respectively.
Surveys completed on the smartwatch application are sent to a cloud database via the phone.
The Cozie platform can also facilitate sending push notifications to users to encourage choices related to comfort or satisfaction.
The push notification logic can be as simple as sending a notification at a predefined time using a time-based job scheduler or sending them as a function of the user's location, physiological variable, or environmental parameters.
Location data are gathered using the GPS module in the watch, while physiological data are queried from the iOS Health Store after obtaining consent from the user.
The GPS location is obtained in real-time each time a survey is completed. 
In addition, the following parameters are logged in the background: heart rate, resting heart rate, blood oxygen saturation, step count, walking distance, stand time, and noise. 
The timing of when these samples are transmitted to the database is inconsistent since iOS manages them and relies on variables such as battery charge level, connectivity, iPhone usage, and other factors. 
More physiological (e.g., sleep quality) and personal data (e.g., race, marital status, education level) can be obtained by querying the Apple Health Store, if need be.
Current development efforts include one-click onboarding, embedding watch surveys without changing the source code, increased reporting frequency of physiological data, and offline data logging.

\subsection{Micro-survey question flow development and experimental deployment}
Figure~\ref{fig:cozie_flow_and_complications} shows an example of an Apple Watch clock face with the Cozie icon (also known as a \emph{complication} when on the watch home screen) and a subset of questions used in this study. 
Complications are small icons that can be added to almost all Apple Watch faces. 
Tapping on the complication will start a watch survey on the Apple watch. 
This feature reduces the effort and the required time for the user to access the survey.

Follow-up questions are possible, i.e., the investigator can determine which question is shown next based on the response given to the previous question. 
The question flow used in this deployment consisted of a total of nine questions that focus on capturing noise distraction and contextual information~\cite{Miller2022-dy}. 
There is no limit on the number of questions that can be included in each question flow, but a range of 4-9 questions is recommended. 

The deployment of Cozie for this paper involved 48 participants between October 2022 and January 2023. 
These participants are primarily staff and students of the National University of Singapore, as they were recruited through an internal website where research studies are posted. 
However, people from the general public who satisfied the criteria were also included. 
The selection criteria were that the participants must be 21 to 65 years old, fluent in English, and willing to wear the watch for 2-4 weeks and answer the micro-surveys at regular intervals.
Participants were required to complete a minimum of 100 surveys in any location, each completed at least one hour apart, and upon completion, were given SG\$100 in compensation. 
More details about the methodology of this deployment can be found in an associated manuscript related to the nudging aspect of the study~\cite{Miller2022-dy}.
This analysis comprises 2,400 micro-survey responses collected by 48 participants who each completed 50 individual surveys.
This analysis is designed to show the platform's potential and initial results and sets the stage for future work.

\section{Results and discussion}
\begin{figure}[b!]%[htbp]
    \centering
    \includegraphics[width=\textwidth]{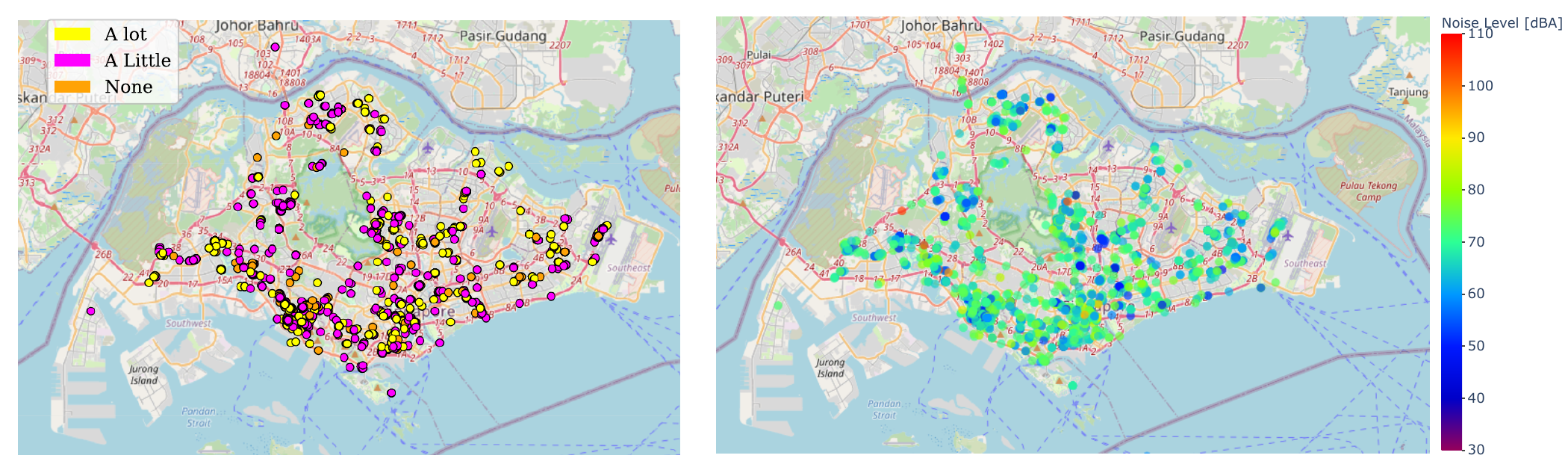} % second figure itself
    \caption{Spatial urban-scale context of the data collected for whether noise distraction is present (left) and the sound levels recorded by the watch sensor (right).}
    \label{fig:orenth_map}
\end{figure}

\begin{figure}%[b!]%[htbp]
    \centering
    \includegraphics[width=0.9\textwidth]{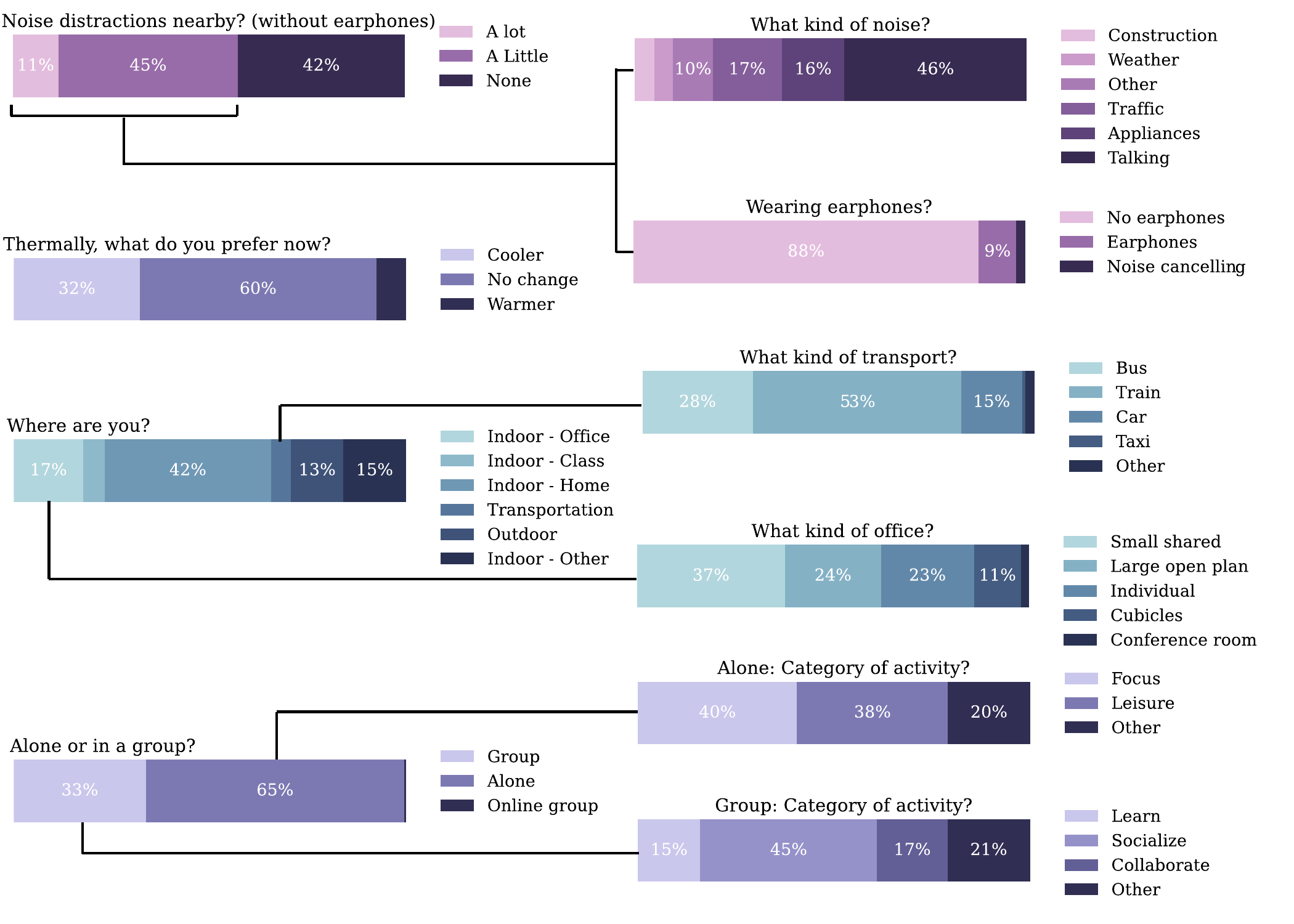} % first figure itself
    \caption{Results breakdown from 48 participants - the bars on the left side contain responses from all 2,400 surveys, while those on the right are data that are only collected based on the responses from previous questions (the chart segment that they are connected to by a line).}    
    \label{fig:orenth_survey_responses}
\end{figure}

Figure~\ref{fig:orenth_map} shows the geographical diversity of where participants provided responses about whether noise distractions were present and the noise levels recorded by the watch sensor in different parts of Singapore. 
Most responses were concentrated in the central region of Singapore and near the National University of Singapore campus.
Figure~\ref{fig:orenth_survey_responses} gives an overview of the data collected from the 48 participants in this deployment. 
The results show that the participants were experiencing noise distraction a slight majority of the time (58\%), with 11\% of the time being significant distractions (\emph{a lot}). 
Almost half (46\%) of the distractions were due to other people \emph{talking} and \emph{traffic} and \emph{appliance} noise being the following two highest distraction sources.
When noise distractions were experienced, the participants most often were not wearing earphones or headphones (88\%).
Thermal preference data showed that the participants preferred \emph{no change} a majority of the time (60\%) and preference for being \emph{cooler} otherwise dominated (31\%) as compared to the preference for being \emph{warmer} (9\%).

In addition to the subjective environmental experiences, data was collected about additional contextual information that could be useful to characterize environmental quality.
The results of these questions showed that participants were in indoor spaces a vast majority of the time, most often at \emph{home} (42\%) or the \emph{office} (17\%). 
When they were in an office, the most common types of spaces were \emph{small shared} (37\%), \emph{large open office} (24\%), and \emph{individual offices} (23\%).
The participants indicated that they were \emph{alone} a majority of the time (65\%) or within a \emph{group} of people (33\%) and only seldom in an \emph{online group} such as a video conferencing meeting.
This data analysis forms the preliminary foundation for future work that dissects the responses according to the characteristics of the geographic location, the individual categories, the demographics from the onboarding survey, and the measured variables of heart rate, step count, and environmental noise levels recorded by the watch.

In this deployment, Cozie enabled gathering feedback from people in real-time in a scalable way.
It also enabled us to couple environmental sensor data, physiological data, and users’ feedback data seamlessly, indicating that the app can be a valuable tool for collecting data in future studies worldwide.
Cozie was built on the previous knowledge that our team gained from developing Cozie for Fitbit~\cite{Jayathissa2019a, Tartarini2022} and a web application to log people's preferences~\cite{Kim2019}.
The Apple platform helps mitigate the limitations of the Fitbit platform, such as the limitation of a maximum of four answers, the lack of choice of a preferred watch face, and the lack of ability to send push notifications to the users.
Cozie also has several applications which extend beyond research settings, including integration into the systems control logic, interventional messages for occupants, and space utilization studies.
Consequently, the platform is open-sourced to reach a broader audience so other researchers, developers, and practitioners can use it.

\section{Conclusions}
\label{sec:conclusions}
This paper outlines the open-source development of the Cozie application for the Apple Watch and iPhone. 
This platform allows people to provide real-time feedback about their surroundings while seamlessly logging physiological, environmental, and behavioral data.
The analysis summarizes 2,400 micro-survey responses from 48 research participants giving high-level insights into how people experience indoor and outdoor spaces in a scalable way. 
These data will be further analyzed in combination with other parts of a more extensive methodology that will be covered in future publications.

\section*{Acknowledgments}
This research was supported by the following Singapore Ministry of Education (MOE) Tier 1 Grants: A-0008305-01-00, A-0008301-01-00, and A-8000139-01-00, and by the Singapore National Research Foundation through a grant to the Berkeley Education Alliance for Research in Singapore (BEARS) for the Singapore-Berkeley Building Efficiency and Sustainability in the Tropics (SinBerBEST) Program. Apple, Apple Watch, iPhone, and iOS are trademarks of Apple Inc., registered in the U.S. and other countries and regions.

% \printbibliography
\section*{References}
\bibliographystyle{elsarticle-num}
\bibliography{references} % .bib

\end{document}